\documentclass[aps,prc,reprint,superscriptaddress,longbibliography]{revtex4-2}
\usepackage{amsmath}
\usepackage{amssymb}
\usepackage{wallpaper}
\usepackage{graphicx}
\usepackage{dcolumn}
\usepackage{bm}
\usepackage{color}
\usepackage{multirow}
\usepackage{CJK}
\usepackage{mathptmx}
\usepackage{tabularx}
\usepackage{tikz-feynman}
\usepackage{booktabs}
\usepackage{array}
\usepackage{ulem}
\usepackage[T1]{fontenc}
\usepackage[colorlinks,linkcolor=blue,anchorcolor=blue,citecolor=blue]{hyperref}
\renewcommand{\emph}[1]{\textit{#1}}

\begin{document}
\begin{CJK*}{UTF8}{gbsn}
\renewcommand\arraystretch{1.2}
\title{
Particle-vibration coupling model with explicit doorway states based on the Skyrme density functional
}

\author{W. L. Lv} %
\address{Frontiers Science Center for Rare isotopes, Lanzhou University, Lanzhou 730000, China}
\address{School of Nuclear Science and Technology, Lanzhou University, Lanzhou 730000, China}
\address{School of Physics and Astronomy, Shanghai Jiao Tong University, 
    Key Laboratory for Particle Astrophysics and Cosmology (MoE), Shanghai 200240, China}
\author{Y. F. Niu} \email{niuyf@sjtu.edu.cn} 
\address{School of Physics and Astronomy, Shanghai Jiao Tong University, 
    Key Laboratory for Particle Astrophysics and Cosmology (MoE), Shanghai 200240, China}
\address{Shanghai Key Laboratory for Particle Physics and Cosmology, Shanghai 200240, China}
\address{Frontiers Science Center for Rare isotopes, Lanzhou University, Lanzhou 730000, China}
\address{School of Nuclear Science and Technology, Lanzhou University, Lanzhou 730000, China}
\author{G. Col\`{o}}
\address{Dipartimento di Fisica, Universit\`{a} degli Studi di Milano, via Celoria 16, I-20133 Milano, Italy}
\address{INFN, Sezione di Milano, via Celoria 16, I-20133 Milano, Italy}

\date{\today}

\begin{abstract}
\noindent
\textbf{Background:}
Modern $\gamma$-ray detectors can achieve energy resolutions of a few keV, 
calling for a detailed microscopic understanding of the strength fragmentation in nuclear excitation spectra. 
Particle-vibration coupling (PVC) plays an important role in describing the spreading of giant resonances.
However, conventional PVC formulations based on an energy-dependent effective Hamiltonian 
in the 1 particle-1 hole (1p1h) space do not explicitly provide 
the individual excited states and their wave functions, thus
limiting the analysis of the microscopic origin of strength fragmentation.

\noindent
\textbf{Purpose:}
The aim of this paper is to develop a PVC model with explicit doorway states,
in which the PVC matrix is directly diagonalised in the space
spanned by 1p1h and
1 particle-1 hole coupled to a phonon
(1p1h$\otimes \nu$) configurations.

\noindent
\textbf{Methods:}
Based on the Skyrme energy density functional SkM*,
we apply the PVC model with explicit doorway states to study the isovector giant dipole resonance (GDR) in $^{208}$Pb.
We analyse the one-body transition densities and 
the microscopic structure of the states carrying large transition strengths.

\noindent
\textbf{Results:}
The one-body transition-density analysis shows that the neutron and proton components 
of the dominant GDR states remain out of phase
after PVC effect is considered,
indicating that the isovector character of the GDR is preserved. The 
microscopic wave-function analysis
reveals that the $2^+$ and $3^-$ phonons play a key role in generating the
fragmentation of the GDR states. The resulting
fragmented states are dominated by 1p1h$\otimes\nu$ configurations, which
account for more than 80\% of the wave-function norm. Despite this strong
reorganization of the wave functions, the collective character of the
$E1$ transition is preserved as reflected by
the coherent contributions of 1p1h configurations.

\noindent
\textbf{Conclusions:}
Therefore, in $^{208}$Pb, the PVC correlations modify the microscopic
wave-function composition and fragmentation pattern of the GDR 
without changing its essential isovector collective character.
\end{abstract}

\maketitle
\end{CJK*}


\section{Introduction}
\label{secIntro}
Giant resonances (GRs) are collective modes of nuclear excitation 
in which a large fraction of the nucleons oscillate coherently~\cite{Bortignon1998,Harakeh2001}. 
GRs have served as a primary probe of the properties of nuclear matter
and the effective nuclear interaction.
In particular, the centroid energies of the isoscalar giant monopole resonance (ISGMR) 
constrain the nuclear incompressibility~\cite{Blaizot1980,Garg2018}, 
while the isovector giant dipole resonance (IVGDR) is closely related to the symmetry energy 
and its density dependence~\cite{RocaMaza2018}. 
The isoscalar giant quadrupole resonance (ISGQR) gives information on the
isoscalar effective mass~\cite{Li2018}.
However, most of the well-established knowledge is still at
the level of global observables, such as the centroid energy
and the exhaustion of the energy-weighted sum rule. 
A more microscopic understanding of the wave functions, 
which is closely related to the damping mechanism of GRs~\cite{Bertsch1983,Drozdz1990}, 
is still highly desirable.

Experimentally, isovector GRs have been extensively studied by
the photonuclear reaction~\cite{Berman1975,Filipescu2024,Zhou2025,Rasulova2025,Fursova2026,Gheorghe2026},
and Coulomb excitation~\cite{Beene1990,Tamii2009,Neveling2011,Rossi2013,Ceruti2017,Brandherm2024}.
Isoscalar GRs are mainly investigated by inelastic
$(\alpha,\alpha')$ scattering~\cite{Youngblood1999,Uchida2004,Li2007,Patel2013,Vandebrouck2014,Krishichayan2015,Gupta2016}, 
$(p,p')$ scattering~\cite{Lewis1972,Wasilewska2022},
and heavy-ion reactions~\cite{Doll1979,Beene1989}.
One of the most important experimental features of GRs is their broad width. 
For instance, the IVGDR typically has a width of about 3--6 MeV~\cite{Plujko2018}. 
This width contains different contributions~\cite{Bertsch1983,Drozdz1990}, 
including Landau damping $\Delta \Gamma$ due to the fragmentation within the 1-particle-1-hole (1p1h) space, 
the escape width associated with direct particle emission $\Gamma^\uparrow$, 
and the spreading width $\Gamma^\downarrow$ generated by the coupling to more complex configurations. 
In medium-heavy and heavy nuclei, the spreading width $\Gamma^\downarrow$
gives a major contribution to the
observed broadening and fragmentation of GR strength distributions.
Theoretically, the key element in understanding the spreading width $\Gamma^\downarrow$
is the description of the coupling of the 1p1h states to 2p2h states.
A microscopic analysis of the GR wave functions is therefore important 
for clarifying how these configurations mix and give rise to the spreading width.

The random phase approximation (RPA) 
has been widely used to describe GRs~\cite{Paar2003,Nakatsukasa2007,Yoshida2008,Bjelcic2020,Chen2026}.
However, it cannot describe the spreading width, 
since only 1p1h configurations are included. 
Various beyond-RPA approaches have therefore
been developed to study GRs.
In the second RPA~\cite{Yannouleas1987,Lacroix2000,Papakonstantinou2010,Gambacurta2010,Vasseur2018,Yang2021,Vale2026,Minato2026}
and in the second Tamm-Dancoff approximation~\cite{Minato2016,Vale2025}, 
2p2h configurations are explicitly included.
In the particle-vibration coupling (PVC) model~\cite{Tselyaev2018,Kamerdzhiev2020,Shen2020,Zhang2022,Litvinova2023,Li2023,Li2024},
the configuration space is extended to include 1p1h configurations coupled to collective phonons, 1p1h$\otimes\nu$.
In the quasiparticle-phonon model~\cite{Soloviev1987,Kamerdzhiev1993,Ponomarev1996,VanGiai1998,Severyukhin2017},
the wave functions are expanded in one-, two-, and three-phonon configurations. 
These complex configurations provide microscopic mechanisms for 
the fragmentation of the strength and for the spreading width of GRs. 
In addition, extended RPA calculations based on 
time-dependent density-matrix theory have shown that 
higher-order correlations in the ground state can also affect the fragmentation of GR strength~\cite{Tohyama2022}.
However, direct analyses of the wave-function composition
remain relatively scarce, with recent exceptions such as Ref.~\cite{Vale2025}.

The PVC model has been successfully used to study both non-charge-exchange~\cite{Li2023,Litvinova2023,Li2024}
and charge-exchange excitations~\cite{Niu2014,Niu2015,Niu2018,Robin2019,Litvinova2020,Liu2024}.
In most previous PVC calculations, the PVC equation is projected onto the
1p1h space, and
the effects of the 1p1h$\otimes\nu$ doorway states are incorporated implicitly through an energy-dependent
complex Hamiltonian~\cite{Colo1994}. This formulation is efficient for
calculating one-body response functions. 
However, since the 1p1h$\otimes\nu$ configurations are not retained explicitly, 
this approach is not well suited for state-by-state studies of
one-body transition densities, microscopic configurations, 
and the collectivity of individual excited states. 
Modern $\gamma$-ray detectors can achieve energy resolutions of a few keV~\cite{Zilges2022}, 
making a detailed theoretical description of individual excited states 
increasingly important for understanding the observed strength fragmentation 
and its underlying microscopic mechanisms.
Furthermore, a detailed comparison between the calculated and experimental strength fragmentation 
may help assess whether the correlations included in the PVC model are sufficient 
or whether higher-order configurations, such as 3p-3h excitations, need to be considered.

This motivates us to develop a 
PVC model with explicit doorway states in which the PVC matrix is diagonalised directly in the
${\rm 1p1h}+{\rm 1p1h}\otimes\nu$ space. 
This formulation gives explicit wave functions for excited states 
and enables a direct analysis of their one-body transition densities
and configuration components. For moderate model-space dimensions, 
direct diagonalisation also provides an efficient way to obtain the complete discrete spectrum 
without evaluating the response function at individual energy points, 
particularly when studying the fine structure of the strength distribution
(see Sec.~\ref{secNume} for a discussion of the computational requirements).
Moreover, the resulting microscopic structure information 
could serve as an input for nuclear reaction calculations~\cite{Andres2001}.

As a first application, we study the IVGDR in $^{208}$Pb 
based on the Skyrme energy density functional.
The paper is organized as follows. In Sec.~\ref{secTheo}, we present the
formalism of the PVC model with explicit doorway states. 
The numerical details and convergence tests are given in Sec.~\ref{secNume}.
In Sec.~\ref{secResu}, we apply the model to the IVGDR in $^{208}$Pb. 
Finally, a summary is given in Sec.~\ref{secSum}.

\section{Formalism}
\label{secTheo}
\subsection{PVC equation in the 1p1h $+$ 1p1h$\otimes \nu$ space}
In the PVC model,
the excited state $|NJM\rangle$ is created by
the operator $\mathcal{O}^{\dag}_{N JM}$,
\begin{equation}
    \begin{aligned}
        &\mathcal{O}^{\dag}_{N JM}
        = \sum_{\rm ph}   X^{(1)N J}_{\rm ph}       A^{\dag}_{\rm ph}(JM) - Y^{(1)N J}_{\rm ph} \tilde{A}_{\rm ph}(JM)  \\
        &\quad +\sum_{{\rm ph} \nu J_s} X^{(2)N J}_{{\rm ph} (J_s)\nu} B^{\dag}_{{\rm ph} (J_s)\nu}(JM)
        -Y^{(2)N J}_{{\rm ph} (J_s)\nu} \tilde{B}_{{\rm ph} (J_s)\nu}(JM),
    \end{aligned}
\end{equation}
where $A^{\dag}_{{\rm ph} }(JM)$ is the creation operator of a 1p1h configuration,
\begin{equation}
    \begin{aligned}
        A^{\dag}_{{\rm ph} }(JM)
        =&~ \sum_{m_{\rm p} m_{\rm h}} (-)^{j_{\rm h}-m_{\rm h}} C_{j_{\rm p} m_{\rm p} j_{\rm h}-m_{\rm h}}^{JM}
        c^{\dag}_{{\rm p} m_{\rm p}} c_{{\rm h} m_{\rm h}},
    \end{aligned}
\end{equation}
and $B^{\dag}_{{\rm ph}(J_s)\nu}(JM)$ is the creation operator of 
a 1p1h state coupled to 1 phonon, that is a
(1p1h $\otimes \nu$) configuration,
\begin{equation}
    \begin{aligned}
        B^{\dag}_{{\rm ph}(J_s)\nu}(JM)
        =  \sum_{M_{s} \mu} C_{J_{s} M_{s} \lambda \mu}^{JM}
        A^{\dag}_{{\rm ph}}(J_{s} M_{s}) Q^{\dag}_{\nu\lambda\mu}.
    \end{aligned}
\end{equation}
Operators $\tilde{A}_{\rm ph}$ and ${\tilde{B}}_{{\rm ph}(J_s)\nu}$ 
are time-reversal of ${A}_{{\rm ph}}$ and ${B}_{{\rm ph}(J_s)\nu}$, 
and are written as
\begin{equation}
    \begin{aligned}
        \tilde{A}_{\rm ph}(JM) =& (-)^{J+M} A_{\rm ph}(J -M),
    \end{aligned}
\end{equation}
and
\begin{equation}
    \begin{aligned}
        {\tilde{B}}_{{\rm ph}(J_s)\nu}(JM) =& (-)^{J+M} B_{{\rm ph}(J_s)\nu}(J-M) .
    \end{aligned}
\end{equation}
The phonon creation operator is
\begin{equation}
    Q^{\dag}_{\nu\lambda\mu}
    = \sum_{{\rm p}_0 {\rm h}_0}  
    X^{\nu\lambda}_{{\rm p}_0 {\rm h}_0}    A^{\dag}_{{\rm p}_0 {\rm h}_0}(\lambda \mu)
    - Y^{\nu\lambda}_{{\rm p}_0 {\rm h}_0} \tilde{A}_{{\rm p}_0 {\rm h}_0}(\lambda \mu) .
\end{equation}

With the equation of motion method~\cite{Schuck2021}, 
the PVC equation can be obtained as
\begin{equation}
\begin{aligned}
  & \left[
    \begin{array}{rrrr}
        A^{(1)}_{1,1'}     &   A^{(2)}_{1,2'}     &   B^{(1)}_{1,1'}     &   B^{(2)}_{1,2'}     \\
        A^{(3)}_{2,1'}     &   A^{(4)}_{2,2'}     &   B^{(3)}_{2,1'}     &   B^{(4)}_{2,2'}     \\
        -B^{(1)\ast}_{1,1'} &  -B^{(2)\ast}_{1,2'} &  -A^{(1)\ast}_{1,1'} &  -A^{(2)\ast}_{1,2'} \\
        -B^{(3)\ast}_{2,1'} &  -B^{(4)\ast}_{2,2'} &  -A^{(3)\ast}_{2,1'} &  -A^{(4)\ast}_{2,2'} \\
    \end{array}
    \right]
    \left[
    \begin{array}{c}
        X^{(1)NJ}_{1'}  \\
        X^{(2)NJ}_{2'}  \\
        Y^{(1)NJ}_{1'}  \\
        Y^{(2)NJ}_{2'}  \\
    \end{array}
    \right]  \\
  & = E_{NJ}
    \left[
    \begin{array}{rrrr}
        G_{1,1'}     &   G_{1,2'}     &   0                  &   0     \\
        G_{2,1'}     &   G_{2,2'}     &   0                  &   0     \\
        0                  &   0                  &   G^{\ast}_{1,1'} &   G^{\ast}_{1,2'} \\
        0                  &   0                  &   G^{\ast}_{2,1'} &   G^{\ast}_{2,2'} \\    
    \end{array}
    \right]
    \left[
    \begin{array}{c}
        X^{(1)NJ}_{1'}  \\
        X^{(2)NJ}_{2'}  \\
        Y^{(1)NJ}_{1'}  \\
        Y^{(2)NJ}_{2'}  \\
    \end{array}
    \right] ,
\label{eq:pvc_full}
\end{aligned}
\end{equation}
where $1$ is a shorthand notation for ph, and $2$ is a shorthand notation for ph$(J_{s})\nu$.
The angular momentum $J_s$ results from the coupling of $j_{\rm p}$ and $j_{\rm h}$.
The normalization of an eigenstate with positive energy $E_{NJ}$ is
\begin{equation}
  1 =  \sum_{\rm ph} \lvert X^{(1)NJ}_{\rm ph} \rvert^2 - \lvert Y^{(1)NJ}_{\rm ph} \rvert^2
     + \sum_{{\rm ph}(J_{s})\nu} \lvert X^{(2)NJ}_{{\rm ph}(J_{s})\nu} \rvert^2 - \lvert Y^{(2)NJ}_{{\rm ph}(J_{s})\nu} \rvert^2 .
     \label{eq:PVC_norm}
\end{equation}

\begin{figure}[tb]
    \centering
    \includegraphics[width=0.3\textwidth]{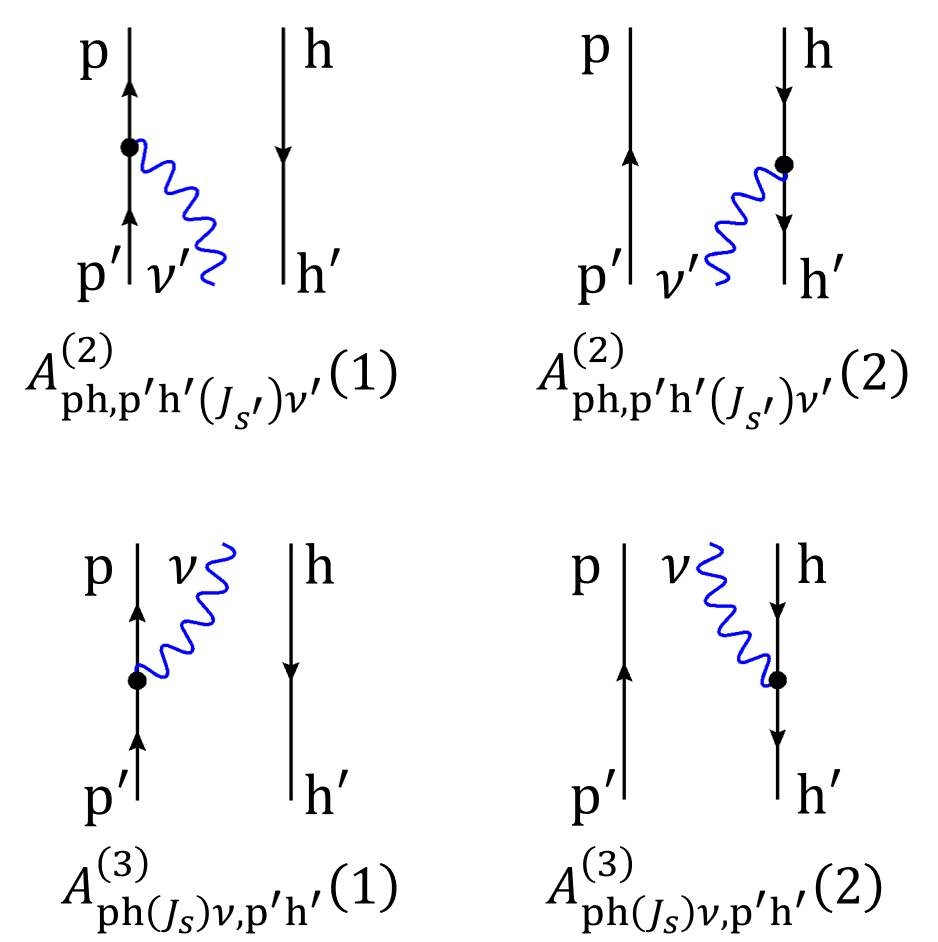}
    \caption{Diagrams for the matrix elements 
             $A^{(2)}_{{\rm ph},{\rm p'h'}(J_{s'})\nu'}$ and $A^{(3)}_{{\rm ph}(J_{s})\nu,{\rm p'h'}}$.
             The solid lines with up and down arrow denote the particle and hole states, respectively.
             The wavy line denotes the phonon.
             The black bullet denotes the PVC vertex.}
    \label{fig:figa23}
\end{figure}

We evaluate the matrices $\mathbf{A}$, $\mathbf{B}$, and $\mathbf{G}$ 
under the quasiboson approximation~\cite{RS1980}.
Matrices $\mathbf{A}^{(1)}$ and $\mathbf{B}^{(1)}$ are same as the ones in RPA model~\cite{Colo2013}. 
Matrices $\mathbf{A}^{(2)}$ and $\mathbf{A}^{(3)}$
arise from the correlation between 1p1h and 1p1h$\otimes \nu$ configurations.
There is a simple relation between them,
\begin{equation}
   \mathbf{A}^{(2)} = \mathbf{A}^{(3)\dag} .
\end{equation}
The matrix elements of $\mathbf{A}^{(2)}$ are evaluated by
\begin{widetext}
\begin{equation}
   A^{(2)}_{{\rm ph},{\rm p'h'}(J_{s'})\nu'}
 = A^{(2)}_{{\rm ph},{\rm p'h'}(J_{s'})\nu'}(1)
  +A^{(2)}_{{\rm ph},{\rm p'h'}(J_{s'})\nu'}(2),
\end{equation}
where
\begin{equation}
    A^{(2)}_{{\rm ph},{\rm p'h'}(J_{s'})\nu'}(1)
    =   \delta_{\rm hh'} 
    (-)^{j_{\rm p}+j_{\rm h'}+\lambda'+J_{s'}}  \hat{J}_{s'} 
    \left\{
    \begin{array}{ccc}
        j_{\rm p'} & j_{\rm h'}     & J_{s'}  \\
        J   & \lambda' & j_{\rm p}
    \end{array}
    \right\} 
    \langle {\rm p} || V || {\rm p'} \nu' \rangle ,
\end{equation}
and
\begin{equation}
  A^{(2)}_{{\rm ph},{\rm p'h'}(J_{s'})\nu'}(2)
=   \delta_{\rm pp'} 
    (-)^{j_{\rm p} + j_{\rm h'} + \lambda' + J} 
    \hat{J}_{s'} 
    \left\{
    \begin{array}{ccc}
        j_{\rm p'}    &  j_{\rm h'}  & J_{s'}  \\
        \lambda' &  J       & j_{\rm h}
    \end{array}
    \right\}  
    \langle ({\rm h})^{-1} || V || ({\rm h'})^{-1} \nu' \rangle.
\end{equation}
with $\hat{J}_{s'}=\sqrt{2J_{s'}+1}$.
We present the diagrams for these matrix elements in Fig.~\ref{fig:figa23}.
The PVC vertices 
$\langle {\rm p} || V || {\rm p'} \nu' \rangle$ and $\langle ({\rm h})^{-1} || V || ({\rm h'})^{-1} \nu' \rangle$
are 
\begin{equation}
\begin{aligned}
    \mathcal{N}^{\lambda'}_{{\rm pp'}}
    \langle {\rm p} || V || {\rm p'} \nu' \rangle
    =&  
    \hat{\lambda}' 
    \sum_{{\rm p}_0 {\rm h}_0} \Big[ X^{\nu'\lambda'}_{{\rm p}_0 {\rm h}_0}  
    \langle [{\rm p}({\rm p}')^{-1}]_{\lambda'}|V|[{\rm p}_0 ({\rm h}_0)^{-1}]_{\lambda'} \rangle 
    \\ & \qquad \qquad \qquad
    -  
    (-)^{j_{{\rm p}_0}+j_{{\rm h}_0}+\lambda'}
    Y^{\nu'\lambda'}_{{\rm p}_0 {\rm h}_0}
    \langle [{\rm p}({\rm p}')^{-1}]_{\lambda'}|V|[{\rm h}_0 ({\rm p}_0)^{-1}]_{\lambda'} \rangle \Big],
\end{aligned}
\end{equation}
and
\begin{equation}
\begin{aligned}
    \mathcal{N}_{\rm hh'}^{\lambda'}
    \langle ({\rm h})^{-1} || V || ({\rm h'})^{-1} \nu' \rangle
    = &
    (-)^{ j_{\rm h} + j_{\rm h'} + \lambda'}
    \hat{\lambda}'
    \sum_{{\rm p}_0 {\rm h}_0} 
    \Big[
    X^{\nu'\lambda'}_{{\rm p}_0 {\rm h}_0}
    \langle [{\rm h'}({\rm h})^{-1}]_{\lambda'}|V|[{\rm p}_0 ({\rm h}_0)^{-1}]_{\lambda'} \rangle
    \\ & \qquad \qquad 
    -
    (-)^{j_{{\rm p}_0}+j_{{\rm h}_0}+\lambda'}
    Y^{\nu'\lambda'}_{{\rm p}_0 {\rm h}_0}
    \langle [{\rm h'}({\rm h})^{-1}]_{\lambda'}|V|[{\rm h}_0 ({\rm p}_0)^{-1}]_{\lambda'} \rangle
    \Big] .
\end{aligned}
\end{equation}
Here $\mathcal{N}_{ab}^{\lambda}$ is the normalization factor,
$\mathcal{N}_{ab}^{\lambda} = \frac{\sqrt{1+ (-)^J \delta_{ab}}}{1+\delta_{ab}}$.
The coupled two-body interaction matrix element is
\begin{equation}
        \langle [{a} {(b)}^{-1}]_J | V | [{c} {(d)}^{-1}]_J \rangle
        =
        \mathcal{N}_{ab}^J \mathcal{N}_{cd}^J
        \sum_{m_a m_b m_c m_d} 
        C_{j_a m_a j_b -m_b}^{J M}
        C_{j_c m_c j_d -m_d}^{J M}
        (-)^{j_b - m_b + j_d - m_d}
        \bar{v}_{adbc} .
\end{equation}
\end{widetext}
For $\mathbf{A}^{(4)}$, we use the diagonal approximation,
\begin{equation}
 A^{(4)}_{{\rm ph}(J_{s})\nu,{\rm p'h'}(J_{s'})\nu'} 
= (\varepsilon_{\rm p}-\varepsilon_{\rm h}+E_{\nu}) 
  \delta_{\rm pp'} \delta_{\rm hh'} \delta_{\nu \nu'} \delta_{J_s J_{s'}} .
\end{equation}
The diagonal approximation has been tested in $^{16}$O 
and found to have a negligible effect on the isoscalar strength distribution of the GDR, 
although it can affect that of the GQR~\cite{Shen2020}.
Because of a mismatch in the total number of creation and annihilation operators,
$\mathbf{B}^{(2)}=\mathbf{B}^{(3)}=\mathbf{B}^{(4)}=\mathbf{0}$.
Owing to the quasiboson approximation, 
the metric matrix $\mathbf{G}=\mathbf{I}$.

For the one-body transition from the ground state 
to excited state $|NJ\rangle$,
the amplitudes $\mathbf{X}^{(2)}$ and $\mathbf{Y}^{(2)}$ do not contribute.
Therefore, the reduced transition strength of a spin-independent operator $\hat{F}_{JM}$ is
\begin{equation}
    B_{NJ}(\hat{F}_{J}) = 
    \sum_{\rm ph} 
    \lvert
    ( X_{\rm ph}^{(1)NJ} + (-)^{J} Y_{\rm ph}^{(1)NJ} )
    \langle j_{\rm p}|| \hat{F}_{J} ||j_{\rm h} \rangle
    \rvert^2 .
\end{equation}
The strength function $S_J(\omega)$ can be obtained by Lorentzian smearing of the discrete strengths $B_{NJ}$,
\begin{equation}
    S_J(E)
=   -\frac{1}{\pi} {\rm Im} \sum_{N} \frac{B_{NJ}}{E - E_{NJ} + \mathrm{i} \eta/2}.
\label{eq:diga_pvc_S}
\end{equation}
where $\eta$ is the Lorentzian smooth parameter.
We focus on the IVGDR in the present paper. 
After removing the contribution from the center of mass,
the isovector dipole transition operator becomes
\begin{equation}
    \hat{F}_{1 M}^{\rm IV}
    = \frac{N}{A} \sum_{i}^{Z} r_{i} Y_{1 M}(\hat{r}_{i})
     -\frac{Z}{A} \sum_{i}^{N} r_{i} Y_{1 M}(\hat{r}_{i}).
\end{equation}
The photoabsorption cross section can be further evaluated by
\begin{equation}
    \sigma(E) = \frac{16 \pi^3}{9} \alpha E S^{\rm IV}_{1} (E),
\end{equation}
where $\alpha$ is the fine-structure constant.
We will also investigate the radial one-body transition density 
of IVGDR, which reads
\begin{equation}
    \delta \rho_{NJ}(r)
    = 
    \hat{J}^{-1}
    \sum_{\rm ph} 
    \frac{ u_{\rm p}(r) u^{\ast}_{\rm h}(r) }{r^2}
    ( X_{\rm ph}^{(1)NJ} + (-)^{J} Y_{\rm ph}^{(1)NJ} )
    \langle j_{\rm p}||Y_{J}||j_{\rm h} \rangle.
\end{equation}
Here $Y_J$ is the spherical harmonic,
and $u_i(r)$ are the single-particle wave functions.

\subsection{PVC equation in the 1p1h space}
If one is only interested in the gross features of the response function,
one can reduce the PVC equation into an energy-dependent
eigen-problem of the dimension of the RPA matrix~\cite{DaProvidencia1965,Colo1994},
\begin{equation}
\begin{aligned}
&    \left[
    \begin{array}{cc}
    A_{1,1'}^{(1)} + W_{1,1'}^{\downarrow}(E) & B_{1,1'}^{(1)} \\
  - B_{1,1'}^{(1)\ast} & -A_{1,1'}^{(1)} - W_{1,1'}^{\downarrow \ast}(-E)
    \end{array}
    \right]
\left[
\begin{array}{c}
    X^{(1)NJ}_{1'}  \\
    Y^{(1)NJ}_{1'}  \\
\end{array}
\right] \\
&\qquad\qquad\qquad\qquad\qquad
=
(E_{NJ} - \mathrm{i} \frac{\Gamma_{NJ}}{2})
\left[
\begin{array}{c}
    X^{(1)NJ}_{1}  \\
    Y^{(1)NJ}_{1}  \\
\end{array}
\right],
\end{aligned}
\label{eq:reduced_pvc}
\end{equation}
where the PVC effects are included through the energy-dependent
spreading term $W_{1,1'}^{\downarrow}(E)$,
\begin{equation}
    W^{\downarrow}_{1,1'}(E)
    =  \sum_{2,2'}
    A^{(2)}_{1,2}
    ( E - A^{(4)}_{2,2'} +\mathrm{i} \eta/2)^{-1}
    A^{(3)}_{2',1'} .
\end{equation}
The spreading term accounts for the coupling to the 1p1h$\otimes\nu$ doorway states 
through an energy-dependent self-energy correction in the 1p1h space. 
In this formulation, the effects of doorway states are incorporated implicitly
rather than treated as explicit degrees of freedom in the eigenvalue problem.
Therefore, we refer to this formulation as the self-energy-based PVC model. 
Hereafter, PVC refers to the model with explicit doorway states developed in the present work,
unless otherwise specified.

Because of the relation
$W^{\downarrow *}_{1,1'}(-E) \neq W^{\downarrow *}_{1,1'}(E)$,
the reduced PVC matrix does not have the standard Bethe-Salpeter/RPA structure.
It represents a non-Hermitian effective eigenvalue problem.
The imaginary parts of the eigenvalues are associated with the spreading widths.
Under the diagonal approximation, the inversion of the large matrix
$A^{(4)}_{2,2'}$ is avoided. 
However, since the 1p1h$\otimes\nu$ configurations are eliminated and enter only
through the energy-dependent spreading term $W^{\downarrow}(E)$, 
the eigenvectors of the reduced equation do not provide direct access to the full wave-function composition. 
Consequently, a state-by-state analysis of the 1p1h$\otimes\nu$ components
and the corresponding one-body transition densities is not straightforward.

The strength function of $\hat{F}_{JM}$ in this case can be derived as
\begin{equation}
   S_J(E) = - \frac{1}{\pi} {\rm Im} 
   \sum_{N} \frac{ \langle 0 || \hat{F}_{J} || NJ \rangle^2 }
                 {E - E_{NJ} + \mathrm{i} (\frac{\Gamma_{NJ}}{2}+\eta/2)}.
\label{eq:reduced_pvc_S}
\end{equation}
To obtain $S_J(E)$, 
Eq.~\eqref{eq:reduced_pvc} is solved at each energy mesh point in the energy region of interest.

\begin{figure*}[t]
    \centering
    \includegraphics[width=1.0\textwidth]{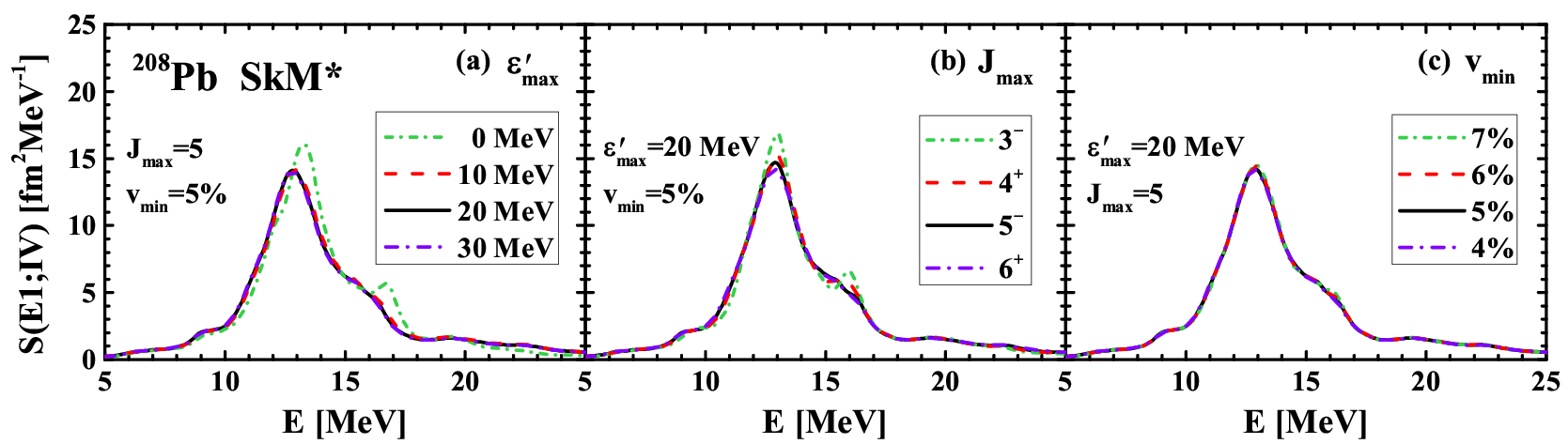}
    \caption{Convergence of the isovector $E1$ strength function in $^{208}$Pb 
        with respect to the three cutoff parameters of the 1p1h$\otimes \nu$ configurations:
        the single-particle energy cutoff $\varepsilon_{\rm max}'$ (a), 
        the maximum phonon angular momentum $J_{\rm max}$ (b), 
        and the phonon collectivity cutoff $v_{\rm min}$ (c).
        See the main text for details of these quantities.
        The width of the Lorentzian function is $\eta=1.0$ MeV.}
    \label{fig:converge}
\end{figure*}

\subsection{Subtraction method}

To avoid the double counting of beyond mean-field many-body correlations,
the subtraction method~\cite{Tselyaev2013} has been adopted in extended RPA models.
In nuclear energy density functional (EDF),
some beyond mean-field many-body correlations are already effectively incorporated 
within the EDF parameterisation.
Accordingly, nuclear static properties can be well described by various EDFs,
including Skyrme, Gogny, and relativistic functionals.
However, an explicit treatment of beyond-mean-field correlations, 
such as particle-vibration coupling, is still necessary to describe dynamical effects, 
including the damping of giant resonances~\cite{Bertsch1983}.
At the same time, we should avoid double counting the 
static correlations already implicitly included in the EDF.
Assuming an exact EDF,
RPA, as the small-amplitude limit of the time-dependent density functional theory,
should provide the exact static response function $R(E=0)$ and the static polarizability $\Pi(E=0)$.
It then follows that these quantities in an extended RPA model
should be consistent with those obtained in the standard RPA.
In practice, the subtraction method amounts to replacing $A^{(1)}_{1,1'}$
in Eqs.~\eqref{eq:pvc_full} and \eqref{eq:reduced_pvc} with $A^{(1)S}_{1,1'}$,
\begin{equation}
   A^{(1)S}_{1,1'}
=  A^{(1)}_{1,1'} - W^{\downarrow}_{1,1'}(0),
\end{equation}
so that the reduced PVC matrix in Eq.~\eqref{eq:reduced_pvc} 
is same as the RPA matrix at $E=0$.

\section{Numerical details}
\label{secNume}

For the ground state,
the Skyrme Hartree-Fock equation is solved in coordinate space 
with $R_{\rm box} =20$ fm and a mesh spacing of 0.1 fm.
The single-particle model space is truncated at $\varepsilon_{\rm max}=100.0$ MeV,
and the angular momentum cutoff is set to $j_{\rm max} = 21/2$.
The SkM* energy density functional is employed throughout this paper.

For the excited states,
since the $2N \times 2N $ PVC matrix is non-Hermitian,
we employ the Ullah-Rowe method~\cite{Ullah1971}, 
which transforms the problem to the diagonalisation of 
two $N \times N$ Hermitian matrices.
The single-particle states entering the 1p1h configurations are 
also truncated at $\varepsilon_{\rm max}=100.0$ MeV.
For the 1p1h$\otimes \nu$ configurations,
three additional cutoffs are introduced to reduce the computational cost, 
namely $\varepsilon_{\rm max}'$, $J_{\rm max}$, and $v_{\rm min}$.
Here, $\varepsilon_{\rm max}'$ is the energy cutoff for the single-particle states 
entering the 1p1h$\otimes \nu$ configurations;
$J_{\rm max}$ is the maximum angular momentum of the phonon with natural parity;
and $v_{\rm min}$ is the cutoff threshold for the phonon selection criterion proposed in Ref.~\cite{Tselyaev2017},
\begin{equation}
    v_{\nu \lambda} = 1
    - \left[\sum_{{\rm p_0 h_0}} ( \lvert X^{\nu \lambda}_{{\rm p_0 h_0}} \rvert^2 - \lvert Y^{\nu \lambda}_{{\rm p_0 h_0}} \rvert^2 )
      ( \varepsilon_{\rm p_0} - \varepsilon_{\rm h_0} ) / E_{\nu \lambda}\right].
\end{equation}
Here, $v_{\nu\lambda}$ serves as a measure of phonon collectivity 
by quantifying the contribution of the residual interaction to the RPA excitation energy. 
In the absence of the residual interaction, the RPA states reduce to unperturbed particle-hole excitations and $v_{\nu\lambda}=0$. 
To select the phonons with non-negligible collectivity, 
only those satisfying $v_{\nu\lambda}>v_{\rm min}$ are retained.
In addition, the phonon energies are restricted to $E_{\nu\lambda}<40$ MeV.

Figure~\ref{fig:converge}(a)--(c) show the corresponding convergence test 
of the isovector $E1$ strength function in $^{208}$Pb. 
Panel~(a) displays the dependence on the single-particle energy cutoff $\varepsilon_{\rm max}'$ 
entering the 1p1h$\otimes \nu$ configuration. 
A pronounced deviation is seen for $\varepsilon_{\rm max}'=0$ MeV, 
whereas the strength distributions obtained with 
$\varepsilon_{\rm max}'=$10, 20, and 30 MeV are already very close to 
each other over the whole energy region. This indicates that the inclusion of 
single-particle states up to $\varepsilon_{\rm max}'=20$ MeV 
in the 1p1h$\otimes \nu$ configuration is sufficient for a stable description. 
Panel~(b) shows the dependence on the phonon angular-momentum cutoff \(J_{\rm max}\). 
While the result with $J_{\rm max}=$3 and 4  still exhibits visible deviations, the curves for 
$J_{\rm max}=$ 5, and 6 are almost same, implying that the $E1$ strength distribution 
is essentially converged at $J_{\rm max}=5$. 
Panel~(c) examines the sensitivity to 
the phonon collectivity cutoff $v_{\rm min}$. 
The results for 
$v_{\rm min}=6$\%, 5\%, and 4\% are very similar.
Therefore, taking both numerical stability 
and computational cost into account, 
we adopt $\varepsilon_{\rm max}'=20$ MeV, $J_{\rm max}=5$, and $v_{\rm min}=5\%$ in the following calculations.
With this numerical condition, the number of 1p1h$\otimes \nu$ configurations is 39790.

The energy-weighted moments $m_1$ listed in Tab.~\ref{tab:ewsr} serve as a check of 
the numerical calculations. For the isovector (IV) dipole strength distribution, 
the RPA results exhaust 99.7\% of the double-commutator sum rule, 
indicating that the model space is large enough.
In the PVC calculations, the $m_1$ values for PVC with subtraction (PVC-S) slightly exceed 100\%, 
whereas PVC without subtraction (PVC-NS) returns to the RPA level. 
This behavior is same as case in SRPA~\cite{Gambacurta2015}.
The inverse energy-weighted moments ($m_{-1}$) serve as a test of the subtraction method.
The subtraction ensures that the static polarizability, $\Pi(E=0)$, 
in the PVC calculation coincides with that of RPA, corresponding to identical $m_{-1}$ values. 
PVC-S reproduces the $m_{-1}$ values of RPA as required,
while PVC-NS slightly deviates due to the double counting of the static correlations.

\begin{table}[h]
    \centering
    \caption{
        Ratios of the energy-weighted moment $m_{1}$ and the inverse energy-weighted moment $m_{-1}$ 
        of the isovector dipole strength distribution in $^{208}$Pb, 
        calculated within RPA, PVC with subtraction (PVC-S), 
        and PVC without subtraction (PVC-NS) using the SkM$^{\ast}$ energy density functional.
        The $m_{1}$ values are normalized to the corresponding double-commutator (D.C.) sum rule. 
        Since no analogous D.C. sum rule is available for \(m_{-1}\), 
        the $m_{-1}$ values are normalized to the RPA result.
    }
    \label{tab:ewsr}
    \renewcommand{\arraystretch}{1.2}
    \begin{tabular}{cccc}
        \hline \hline
        Model                                  & RPA    & PVC-S   & PVC-NS  \\  \hline
        $m_{1}(\rm IV)/m_{1}^{\rm D.C.}(IV)$   & 99.7\% & 103.2\% & 99.7\%  \\
        $m_{-1}(\rm IV)/m_{-1}^{\rm RPA}(IV)$  &100.0\% & 100.0\% &102.5\%  \\
        \hline \hline
    \end{tabular}
\end{table}

\begin{figure}[tb]
    \centering
    \includegraphics[width=0.48\textwidth]{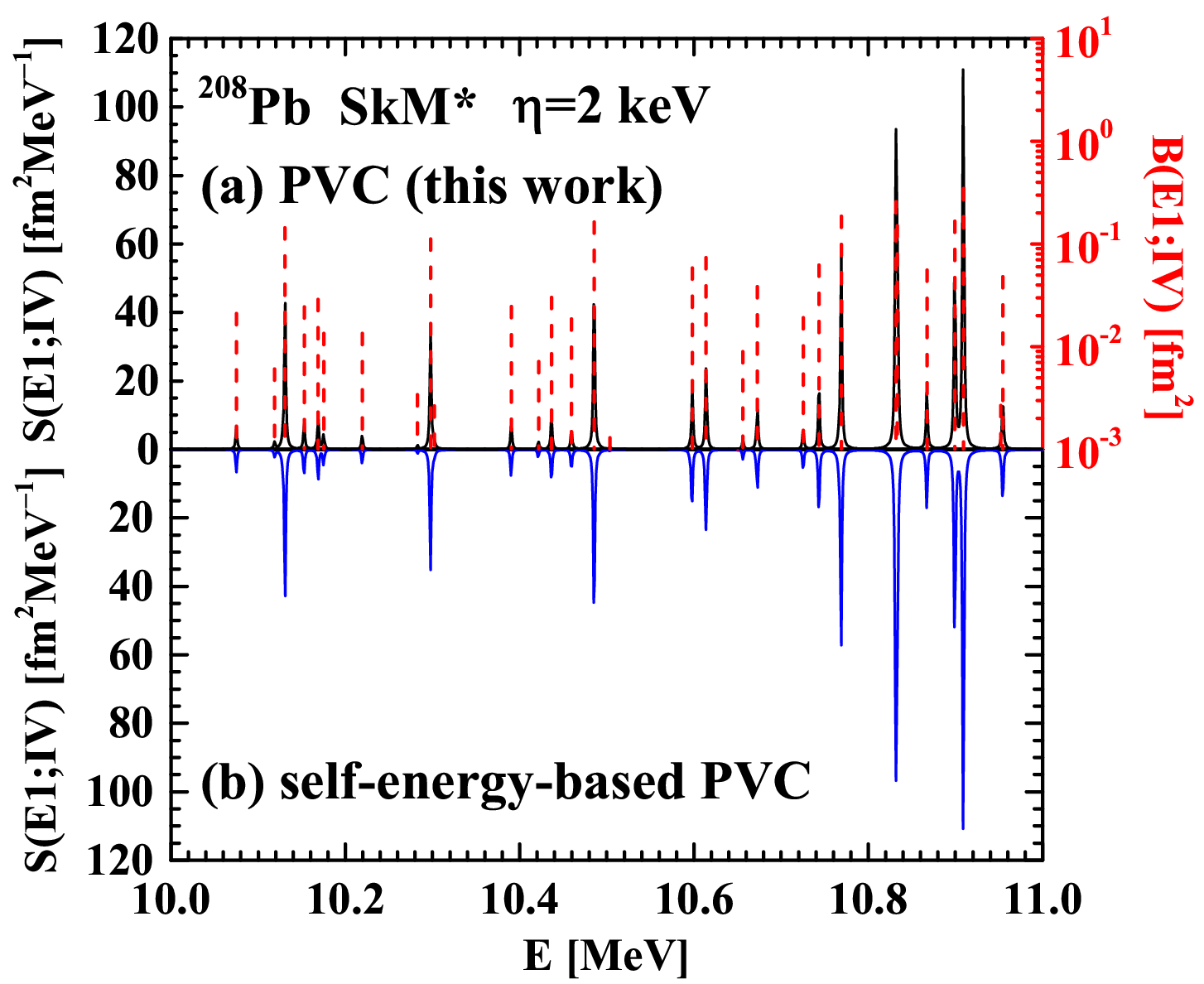}
    \caption{Comparison of the strength functions calculated with the PVC model
        with explicit doorway states (a) and the self-energy-based PVC model (b),
        using the SkM* interaction and a Lorentzian width of $\eta=2$ keV.
        The discrete transition strengths are also shown by red dashed
        vertical lines in panel (a).}
    \label{fig:sth_diag}
\end{figure}

In Fig.~\ref{fig:sth_diag}, 
we compare the two PVC formulations using a very small
smearing parameter, $\eta/2=1$ keV, in the energy interval of 10--11 MeV.
We find a one-to-one correspondence between the transition strengths
obtained from the two approaches, confirming the consistency between
the formulation based on explicit doorway states and that based on the self-energy. 
The calculations were 
performed on AMD EPYC 9654 processors using 100 OpenMP threads.
After preparing the PVC vertices, under the present numerical conditions,
the self-energy-based PVC requires about 600 CPU hours to calculate 
the response function at 1000 energy points with a step of 1 keV, 
whereas diagonalising the PVC equation in the full model space requires 
only about 80 CPU hours to obtain the complete discrete spectrum. 
However, the memory requirement increases from approximately 5 GB to 60 GB. 
These results indicate that the latter approach is computationally 
more efficient for high-resolution studies, at the expense of increased memory usage.

\section{Results and discussions}
\label{secResu}

\begin{figure}[tb]
    \centering
    \includegraphics[width=0.48\textwidth]{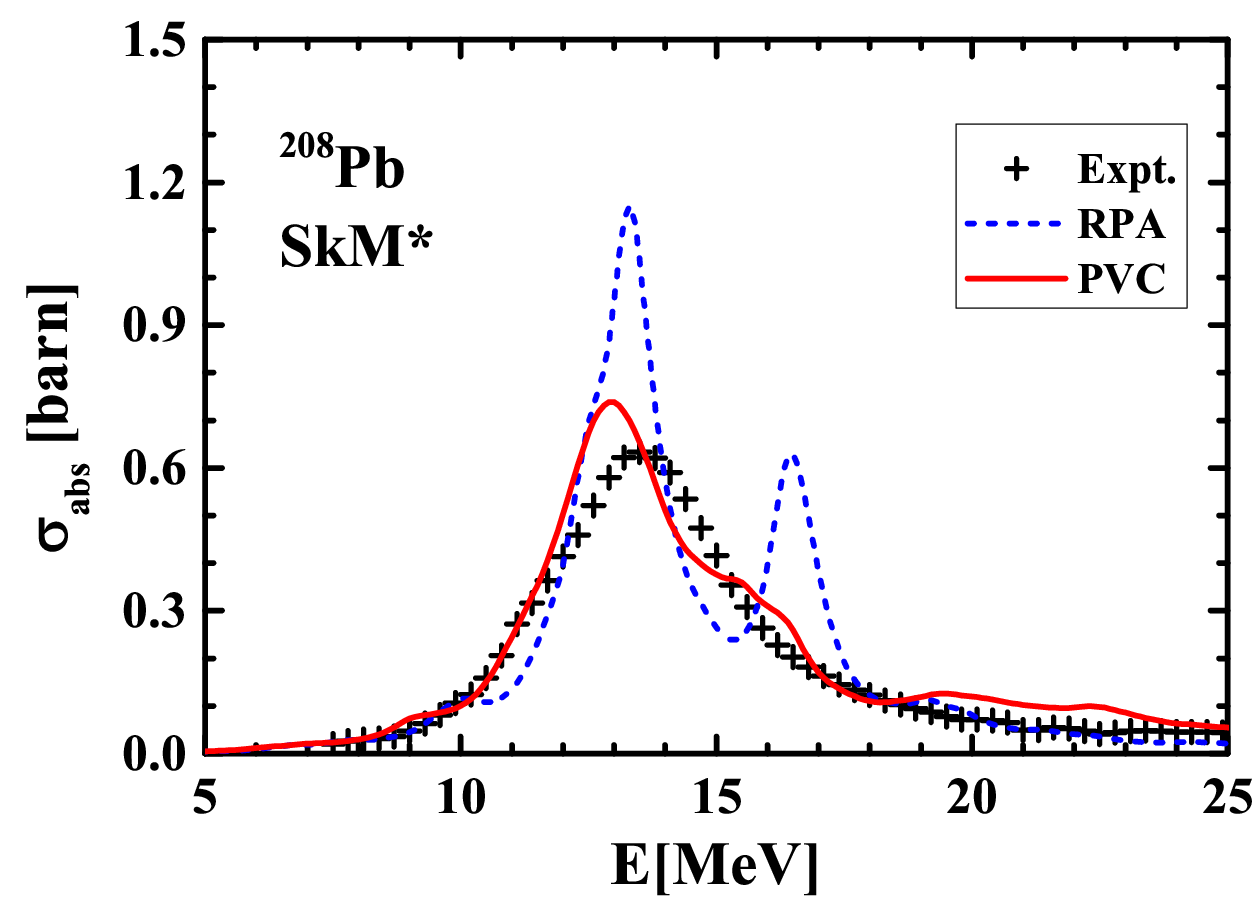}
    \caption{Electric dipole photoabsorption cross section in $^{208}$Pb calculated 
        within the RPA model (dashed line) and PVC model (solid line), 
        using the SkM* interaction folded with a Lorentzian of width $\eta=1$ MeV.
        Experimental data~\cite{Varlamov2013} are shown by the black crosses.}
    \label{fig:xs_B}
\end{figure}

\begin{figure}[tb]
    \centering
    \includegraphics[width=0.48\textwidth]{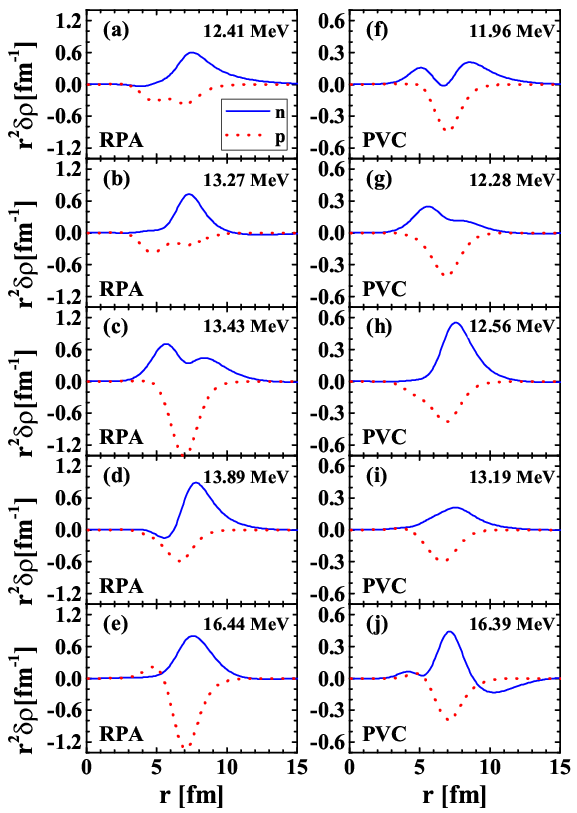}
    \caption{Neutron (blue solid line) and proton (red dashed line) one-body transition densities of
        the states carrying the largest isovector $E1$ strengths. 
        Panels (a)--(e) show the RPA results, while panels (f)--(j) show the PVC results. 
        The excitation energy of each state is indicated in the corresponding panel.}
    \label{fig:TD_GDR}
\end{figure}

\begin{figure}[tb]
    \centering
    \includegraphics[width=0.48\textwidth]{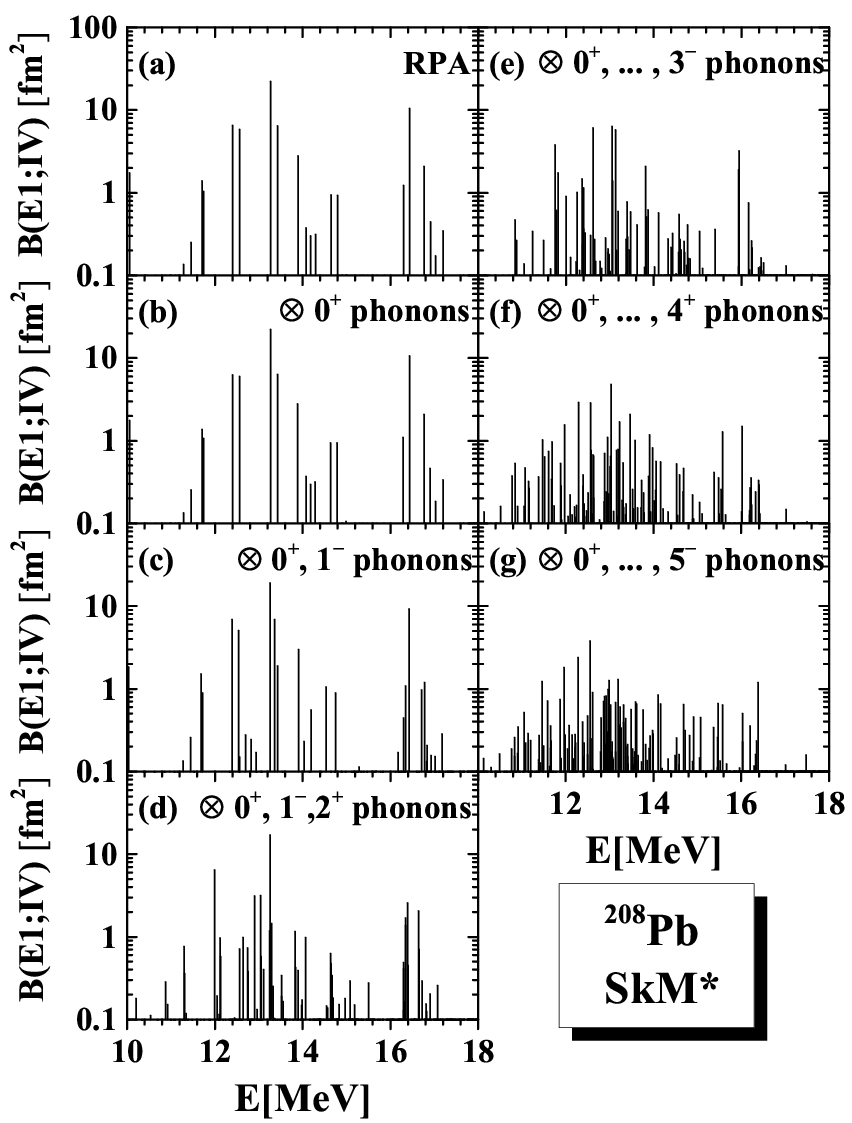}
    \caption{
        Discrete isovector dipole transition strength $B(E1;{\rm IV})$ in
        $^{208}$Pb calculated with the SkM* functional.
        The RPA result is shown in panel (a). 
        Panels (b)--(g) correspond to PVC calculations 
        including successively the $0^+$, $1^-$, $2^+$,
        $3^-$, $4^+$, and $5^-$ phonons, respectively.
    }
    \label{fig:dis_B}
\end{figure}

To assess the PVC model with explicit doorway states, 
we compare the calculated electric dipole photoabsorption cross section 
in $^{208}$Pb with experiment
and benchmark it against the self-energy-based PVC formulation.
Figure~\ref{fig:xs_B} shows the results of RPA and PVC models.
The same Lorentzian width $\eta=1.0$ MeV is adopted for all calculations.
The RPA calculation produces two pronounced peaks at 
$E=13.27$ and $16.44$ MeV. 
After including the PVC correlations, the strength distribution is
redistributed over a larger number of states. 
In particular, the first prominent peak is shifted downward by about
1 MeV, while the pronounced structure around $16$ MeV becomes strongly fragmented. 
As a result, the calculated photoabsorption cross section becomes smoother
and broader, showing improved agreement with the experimental data.
This behavior reflects the spreading width generated by the coupling of
1p1h configurations to 1p1h$\otimes$phonon doorway states.

The PVC effect redistributes the collective GDR strength of the RPA state
over many fragmented states. 
It is therefore important to examine whether these
fragments retain the isovector character of the original RPA excitation.
To address this point, we plot in Fig.~\ref{fig:TD_GDR} the neutron and proton one-body transition densities of the states
carrying the largest isovector $E1$ strengths in the GDR region of $^{208}$Pb. 
These states may therefore be regarded as representative of the dominant dipole response in this energy interval.
For the RPA states [panels (a)--(e)], the neutron and proton transition densities are predominantly 
out of phase in all cases, confirming their isovector character.
For the PVC states [panels (f)--(j)], the transition densities 
are qualitatively similar to those obtained in the RPA,
but exhibit smaller amplitudes due to the fragmentation of the transition strength among several PVC states.
These results indicate that the PVC coupling fragments the strength without qualitatively changing
the isospin character of the underlying GDR excitation.

To examine how the PVC correlations modify the discrete dipole spectrum,
we analyse the evolution of the isovector transition strength $B(E1;{\rm IV})$
with the successive inclusion of phonon configurations, as shown in Fig.~\ref{fig:dis_B}.
In the RPA calculation [panel (a)], 
the dipole strength is mainly concentrated in several pronounced transitions.
The inclusion of the $0^+$ [panel (b)] and $1^-$ [panel (c)] phonons 
only slightly modifies the discrete strength distribution, 
with the main peaks remaining almost unchanged.
After including the $2^+$ phonons [panel (d)], 
the main GDR peak is slightly fragmented. 
At the same time, the high-energy peak around 16.4 MeV, which is prominent in the RPA calculation,
becomes strongly fragmented, with its strength spread over several neighboring states.
With the further inclusion of higher-multipolarity phonons from
$3^-$ to $5^-$ [panels (e)--(g)], the fragmentation pattern becomes
progressively richer, with an increasing number of weak transitions
appearing in the GDR region and a gradual shift of the main strength
distribution toward lower excitation energies.

\begin{figure}[tb]
    \centering
    \includegraphics[width=0.48\textwidth]{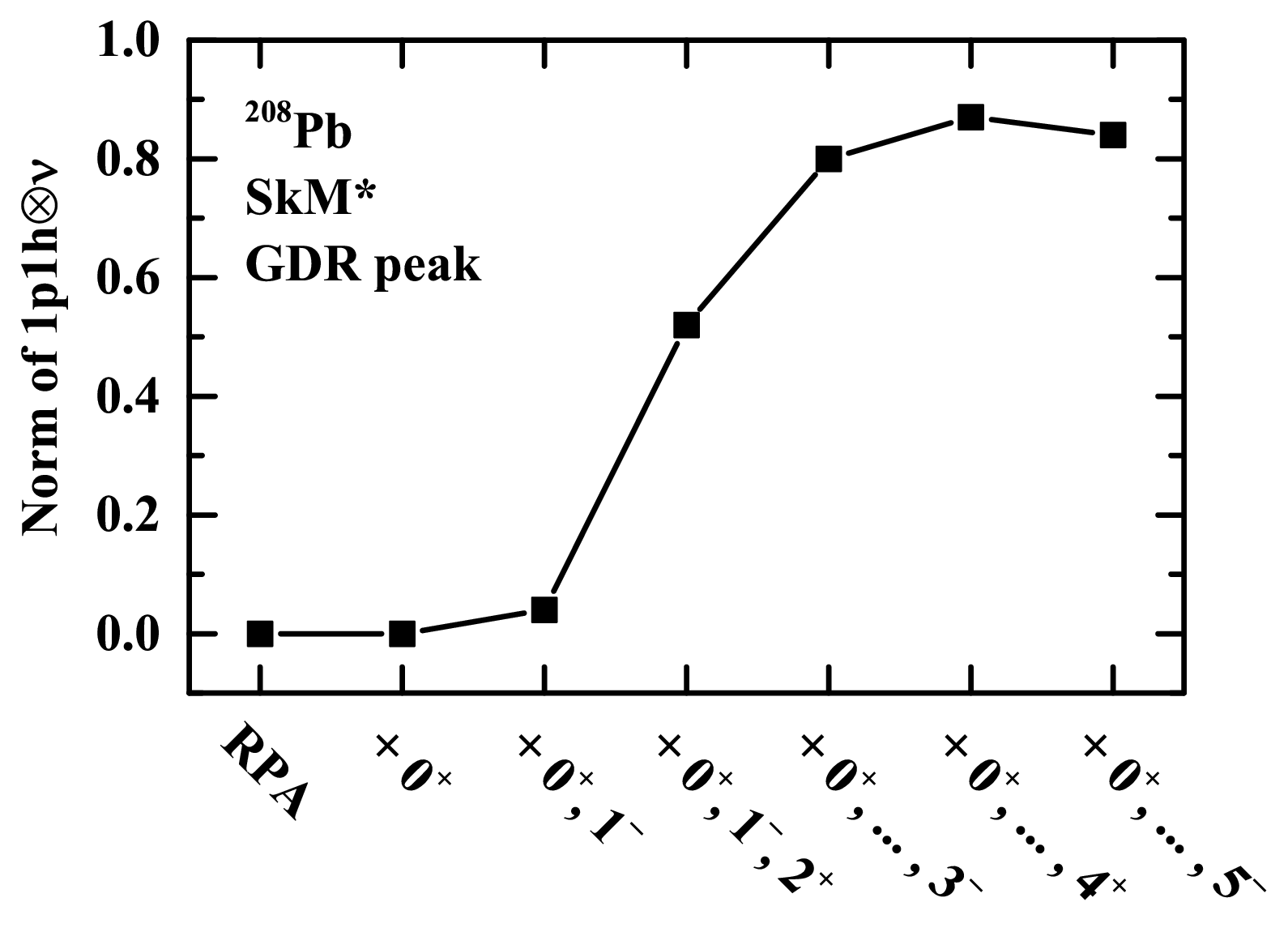}
    \caption{Evolution of the norm of the 1p1h$\otimes\nu$ components in the wave
        function of the main GDR peak around 13 MeV with the successive inclusion of
        $0^+$, $1^-$, $2^+$, $3^-$, $4^+$, and $5^-$ phonons in the PVC
        calculation.}
    \label{fig:norm_pho}
\end{figure}

With increasing phonon space, the corresponding changes in the microscopic
wave function composition provide further insight into the PVC correlations. 
The norm contribution of a configuration $k$ is defined as
\begin{equation}
    w_k=|X_k|^2-|Y_k|^2 ,
\end{equation}
where $k$ denotes either a 1p1h or a 1p1h$\otimes\nu$ configuration.
Figure~\ref{fig:norm_pho} shows the evolution of the
1p1h$\otimes\nu$ component in the wave function of the main GDR state
around 13 MeV with the successive inclusion of phonon configurations.
With the inclusion of the $0^+$ and $1^-$ phonons, the
1p1h$\otimes\nu$ admixture remains very small, indicating that these
phonon configurations only weakly modify the RPA state. A pronounced
increase appears after including the $2^+$ phonons, where the
1p1h$\otimes\nu$ component reaches about 50\%. The subsequent inclusion
of the $3^-$ phonons further enhances this component to about 80\%.
After the inclusion
of higher-multipolarity phonons of $4^+$ and $5^-$, the
1p1h$\otimes\nu$ norm becomes nearly saturated around 85\%. This
indicates that the major restructuring of the main GDR state has already
been established by the $2^+$ and $3^-$ phonons. 
The higher-multipolarity phonons mainly introduce additional mixing within
the already strongly mixed state, which modifies the detailed
fragmentation pattern and fine structure of the dipole response
while only slightly changing the relative ratio
between the 1p1h and 1p1h$\otimes\nu$ components.
The decrease in the 1p1h component of the main GDR state is accompanied 
by a reduction in its transition strength. 
However, this reduction is partially compensated by the appearance of 
many additional states carrying weaker transition strengths as illustrated in Fig.~\ref{fig:dis_B}.

\begin{table*}[t]
    \centering
    \caption{
Configurations of the GDR peak at 13.27 MeV in the RPA model and at 12.56 MeV in the PVC model for $^{208}$Pb
based on the SkM* functional. For each configuration, the contribution $w$ to the norm 
and the partial contribution to the reduced transition amplitude, $b_{\rm ph}$, are shown. 
The 1p1h$\otimes\nu$ configurations do not contribute directly to the
one-body transition operator, and therefore no $b_{\rm ph}$ values are
given for these components. 
}
    \label{tab:pvc_gdr_conf}
    \begingroup
    \setlength{\tabcolsep}{4pt}
    \renewcommand{\arraystretch}{1.0}
    \begin{tabular}{lrr@{\hspace{0.8cm}}lrr}
        \hline\hline
        RPA & ~ &~ & PVC & ~ &~ \\ \hline
        Configuration & \multicolumn{1}{c}{$w$} & {$b_{\rm ph}$ (e$\cdot$fm)} &
        Configuration & \multicolumn{1}{c}{$w$} & {$b_{\rm ph}$ (e$\cdot$fm)} \\
        
        $\nu 2h_{11/2}, \nu 1i_{13/2}$ & 0.254 & $-0.454$ &
        $\pi 2f_{7/2}, \pi 1g_{9/2}$ & 0.046 & $-0.274$ \\
        
        $\pi 2f_{5/2}, \pi 1g_{7/2}$ & 0.149 & 0.417 &
        $\nu 3g_{7/2}, \nu 2f_{5/2}$ & 0.022 & $0.023$ \\
        
        $\nu 1i_{11/2}, \nu 1h_{9/2}$ & 0.123 & 0.824 &
        $\pi 1h_{9/2}, \pi 1g_{7/2}$ & 0.019 & $0.430$ \\
        
        $\nu 1j_{15/2}, \nu 1i_{13/2}$ & 0.065 & 0.754 &
        $[\pi 2f_{7/2},\pi 2d_{5/2}]_{1^-}\otimes 2^+_{1}$ & 0.077 & \\
        
        $\pi 1i_{13/2}, \pi 1h_{11/2}$ & 0.065 & 1.023 &
        $[\nu 1i_{11/2},\nu 2f_{5/2}]_{4^-}\otimes 4^+_{1}$ & 0.076 & \\
        
        $\nu 2g_{7/2}, \nu 1h_{9/2}$ & 0.051 & 0.178 &
        $[\pi 1i_{13/2},\pi 1h_{11/2}]_{1^-}\otimes 2^+_{1}$ & 0.059 & \\
        
        $\pi 1h_{9/2}, \pi 1g_{7/2}$ & 0.036 & 0.575 &
        $[\nu 1i_{11/2},\nu 2f_{5/2}]_{3^-}\otimes 4^+_{1}$ & 0.058 & \\
        
        $\nu 6s_{1/2}, \nu 3p_{1/2}$ & 0.031 & $-0.014$ &
        $[\nu 2g_{9/2},\nu 1h_{9/2}]_{4^-}\otimes 4^+_{1}$ & 0.056 & \\
        
        $\nu 2g_{9/2}, \nu 1h_{11/2}$ & 0.025 & $-0.150$ &
        $[\pi 2f_{5/2},\pi 1h_{11/2}]_{3^+}\otimes 3^-_{1}$ & 0.046 & \\
        
        $\nu 2g_{9/2}, \nu 2f_{7/2}$ & 0.023 & 0.302 &
        $[\nu 1j_{15/2},\nu 1h_{9/2}]_{3^+}\otimes 3^-_{1}$ & 0.045 & \\
        
        $\pi 2f_{7/2}, \pi 2d_{5/2}$ & 0.017 & 0.318 &
        $[\nu 3d_{5/2},\nu 2f_{5/2}]_{3^-}\otimes 4^+_{1}$ & 0.031 & \\
        
        & & &
        $[\pi 2f_{5/2},\pi 1h_{11/2}]_{4^+}\otimes 3^-_{1}$ & 0.025 & \\
        
        & & &
        $[\nu 1i_{11/2},\nu 2f_{5/2}]_{5^-}\otimes 4^+_{1}$ & 0.024 & \\
        
        & & &
        $[\pi 1h_{9/2},\pi 1h_{11/2}]_{2^+}\otimes 1^-_{1}$ & 0.023 & \\
        
        & & &
        $[\pi 2f_{7/2},\pi 1h_{11/2}]_{3^+}\otimes 3^-_{5}$ & 0.022 & \\
        
        & & &
        $[\pi 2f_{7/2},\pi 2d_{5/2}]_{2^-}\otimes 2^+_{1}$ & 0.018 & \\
        
        & & &
        $[\nu 1j_{15/2},\nu 1i_{13/2}]_{1^-}\otimes 2^+_{1}$ & 0.017 & \\
        
        & & &
        $[\nu 1j_{15/2},\nu 1h_{9/2}]_{4^+}\otimes 3^-_{1}$ & 0.017 & \\
        
        & & &
        $[\pi 2f_{7/2},\pi 1h_{11/2}]_{4^+}\otimes 3^-_{5}$ & 0.015 & \\
        
        & & &
        $[\pi 2f_{7/2},\pi 1h_{11/2}]_{2^+}\otimes 3^-_{5}$ & 0.013 & \\
        
        & & &
        $[\pi 1h_{9/2},\pi 2d_{5/2}]_{2^-}\otimes 2^+_{1}$ & 0.013 & \\
        
        \hline
        Number of 1p1h configurations &~ &~&
        Number of 1p1h configurations &~ &~\\
        
        11 with $w>0.01$ & 0.839 & 3.774 &
        3 with $w>0.01$ & 0.087 & $0.180$ \\
        
        33 with $w\in[0.001,0.01]$ & 0.146 & 0.938 &
        18 with $w\in[0.001,0.01]$ & 0.058 & $1.580$ \\
        
        1087 with $w<0.001$ & 0.015 & $-0.004$ &
        1110 with $w<0.001$ & 0.014 & $0.189$ \\

        Sum of 1p1h configurations & 1.000 & $4.708$ &
        Sum of 1p1h configurations & 0.159 & $1.949$  \\

        & & &
        Number of 1p1h$\otimes\nu$ configurations&~&~ \\
        
        & & &
        18 with $w>0.01$ & 0.637 & \\
        
        & & &
        48 with $w\in[0.001,0.01]$ & 0.157 & \\
        
        & & &
        39724 with $w<0.001$ & 0.047 & \\

        & & &
        Sum of 1p1h$\otimes \nu$ configurations & 0.841 &  \\        
        \hline\hline
    \end{tabular}
    \endgroup
\end{table*}

To obtain a more detailed microscopic picture of the main GDR peak, 
we analyse the contributions of individual configurations to the wave-function norm. 
In addition, the collectivity of the transition 
is assessed through the partial contributions of individual particle-hole configurations 
to the reduced transition amplitude, defined as
\begin{equation}
    b_{\rm ph} = 
    ( X_{\rm ph}^{(1)NJ} + (-)^{J} Y_{\rm ph}^{(1)NJ} )
    \langle j_{\rm p}|| \hat{F}_{J} ||j_{\rm h} \rangle .
\end{equation}
Table~\ref{tab:pvc_gdr_conf} summarizes the microscopic structure of the main GDR peak at 13.27 MeV in the RPA 
and the corresponding PVC state at 12.56 MeV. In the RPA, the state is entirely composed of 1p1h configurations. 
The 11 configurations with $w>0.01$ account for 83.9\% of the total norm and contribute 3.774 e$\cdot$fm to the reduced transition amplitude. 
The remaining configurations contribute an additional 0.934 e$\cdot$fm, despite their relatively small individual weights. 
This indicates a coherent superposition of particle-hole excitations, 
reflecting the collective character of the GDR.
The inclusion of particle-vibration coupling substantially changes the composition of the state. 
The total 1p1h weight decreases to 0.159, while the 1p1h$\otimes\nu$ configurations account for 0.841 of the norm. 
Among the latter, the 18 configurations with $w>0.01$ contribute 0.637,
whereas the 39724 configurations with $w<0.001$ contribute only 0.047. 
Thus, although the PVC space contains a large number of complex configurations, 
most of their total weight is concentrated in a relatively small number of dominant components.
Despite the strong reduction of the 1p1h norm, the PVC state retains coherent contributions to the $E1$ transition amplitude. 
The 1p1h configurations with $w>0.01$, $0.001\leq w\leq0.01$, and $w<0.001$ contribute $0.180$, $1.580$, and $0.189$ e$\cdot$fm, respectively, 
yielding a total reduced transition amplitude of $1.949$ e$\cdot$fm. 
Notably, the configurations with intermediate weights ($0.001\leq w\leq0.01$) make the largest contribution to the transition amplitude, 
despite their relatively small contribution to the wave-function norm. 
Thus, although PVC redistributes most of the norm into 1p1h$\otimes\nu$ configurations, 
the coherent contribution of the 1p1h sector to the $E1$ response remains evident.

\begin{figure}[tb]
    \centering
    \includegraphics[width=0.48\textwidth]{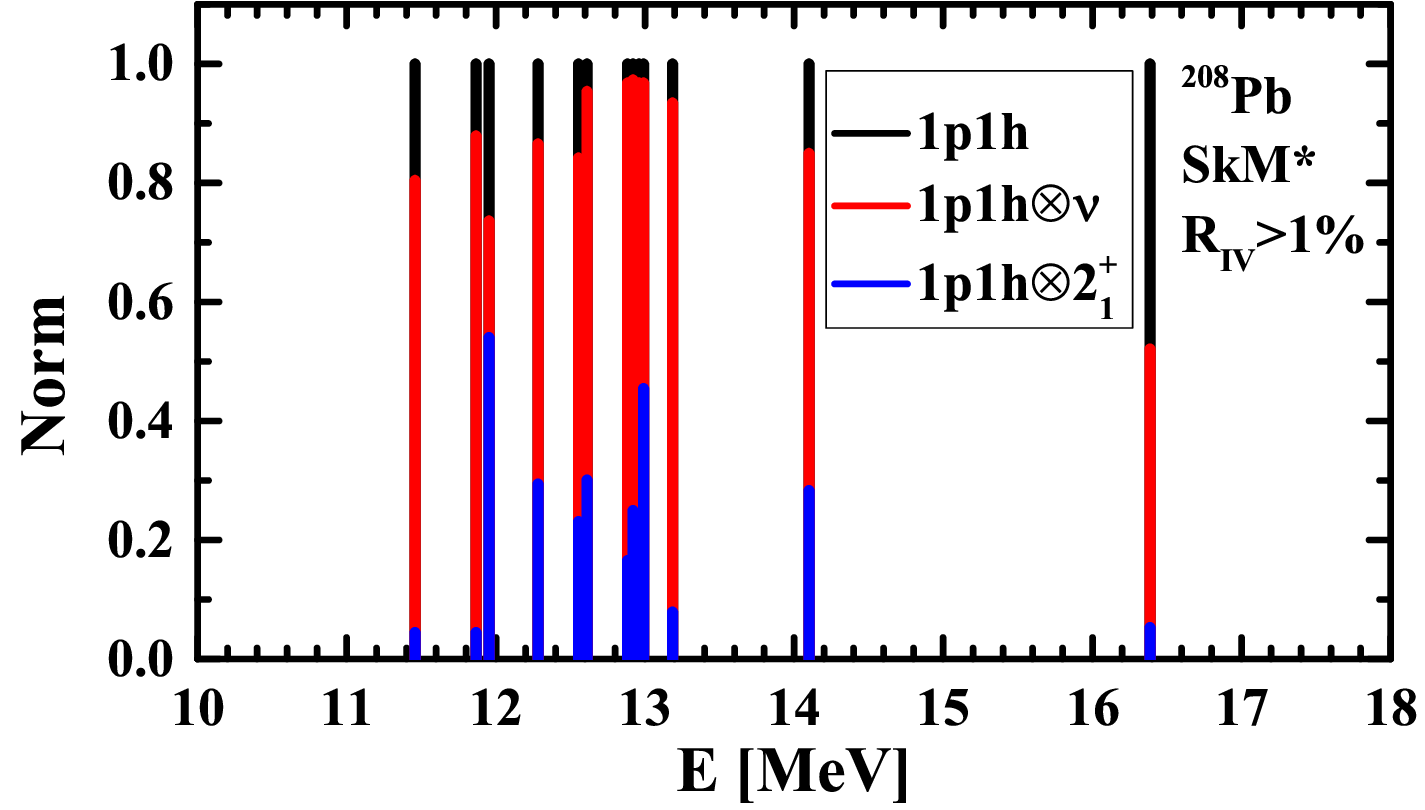}
    \caption{Composition of the excited-state wave functions in $^{208}$Pb calculated with the PVC model
    in the energy interval 10--18 MeV. The black bars denote the 1p1h contribution, and the red bars the 1p1h$\otimes \nu$ contribution.
    The latter includes 1p1h$\otimes 2_{1}^{+}$ component (blue).
    Panel (a) shows the PVC results, while panel (b) shows the corresponding results obtained with the $B$ matrix switched off. 
    Only states carrying more than 1\% of the IV transition strengths are shown ($R_{\rm IV}>1\%$).}
    \label{fig:comp_GDR}
\end{figure}

Finally, we extend the configuration analysis to the fragmented dipole states 
carrying more than 1\% of the total isovector $E1$ strength in the GDR region.
Figure~\ref{fig:comp_GDR} shows that the strong mixing observed for the
main GDR state is also present throughout the GDR region. 
For all prominent dipole states, the 1p1h$\otimes\nu$ components dominate the wave
functions, contributing more than 80\% of the total norm, 
while the 1p1h components provide only a minor contribution. 
Among the phonon-coupled configurations, 
the 1p1h$\otimes 2_1^+$ component provides a remarkable contribution.
This is in line with its important role found in the above analysis. 
Our previous calculations showed a non-negligible $\gamma$-decay 
branching ratio of the GDR to the $2_1^+$ state~\cite{Lv2021}, 
suggesting the presence of a 1p1h$\otimes 2_1^+$ admixture in the GDR wave function.
The present wave-function analysis provides a direct quantification of this component.

\section{Summary}
\label{secSum}
In this work, we developed the PVC model with explicit doorway states
based on the Skyrme energy density functional. 
In the conventional self-energy-based PVC approach, 
the full PVC equation is projected onto the 1p1h subspace, 
and the effects of the 1p1h$\otimes\nu$ doorway states 
are incorporated implicitly through an energy-dependent effective Hamiltonian.
The resulting eigenvalue problem has the same dimension as the RPA matrix 
and is suitable for calculating one-body response functions.
In the present work, instead, we diagonalise the PVC matrix directly in
the 1p1h$+$1p1h$\otimes\nu$ space. 
This formulation gives explicit excited-state wave functions 
and makes it possible to analyse one-body transition densities 
and microscopic configurations of individual states.

The model was applied to the isovector giant dipole resonance in
$^{208}$Pb with the SkM* interaction.
The calculated photoabsorption cross section shows that 
the RPA strength is concentrated in two pronounced peaks. 
When the PVC effect is included, 
the GDR strength is shifted downward and redistributed over many more states. 
As a result, the PVC calculation produces a broader and smoother distribution, 
in better agreement with experiment. 
The transition-density analysis shows that 
the neutron and proton one-body transition densities remain mainly out of phase in the surface region. 
Thus, the PVC effect shifts the GDR strength to lower energy and fragments it, 
but does not qualitatively change the isovector character of the GDR.

The microscopic effects of the phonon coupling were further investigated
through the evolution of the discrete $B(E1;{\rm IV})$ distribution and
the wave-function composition. 
The inclusion of the $2^+$ and $3^-$ phonons plays a key role 
in restructuring the GDR wave function. 
Higher-multipolarity phonons mainly modify the detailed fragmentation pattern and fine structure. 
The wave-function analysis shows that the fragmented GDR
states are dominated by 1p1h$\otimes\nu$ components, 
which carry more than 80\% of the norm in the GDR region. 
Meanwhile, the coherent contribution of 1p1h configurations
to the transition amplitude indicates that the collectivity of the GDR is maintained.

\section*{Acknowledgements}
This work was supported by 
the National Natural Science Foundation of China under grant Nos. 12405135, 12447168, 12075104, 12447106, 
the ``Young Scientist Scheme'' of National Key Research and Development (R\&D) Program under grant No. 2021YFA1601500,
the Science and Technology Innovation Leading Talent Project of Gansu Province (25RCKA025), 
the Lingchuang Research Project of China National Nuclear Corporation (CNNC-LCKY-2024-082), 
and the Fundamental Research Funds for the Central Universities (lzujbky-2023-stlt01).

\section*{Data availability}
The data that support the findings of this article are not
publicly available. The data are available from the authors
upon reasonable request.

%

\end{document}